\documentclass{article}

\usepackage{arxiv}

\usepackage[utf8]{inputenc} % allow utf-8 input
\usepackage[T1]{fontenc}    % use 8-bit T1 fonts
\usepackage{hyperref}       % hyperlinks
\usepackage{url}            % simple URL typesetting
\usepackage{booktabs}       % professional-quality tables
\usepackage{amsfonts}       % blackboard math symbols
\usepackage{nicefrac}       % compact symbols for 1/2, etc.
\usepackage{microtype}      % microtypography
\usepackage{cleveref}       % smart cross-referencing
\usepackage{lipsum}         % Can be removed after putting your text content
\usepackage{graphicx}
\usepackage[round]{natbib}
\usepackage{doi}

\usepackage{array}			% added DK for raggedright

\title{The Buy-or-Build Decision, Revisited: How Agentic AI Changes the Economics of Enterprise Software}

\newif\ifuniqueAffiliation
\uniqueAffiliationtrue

\ifuniqueAffiliation % Standard variant of author block
\author{ \href{https://orcid.org/0000-0001-8322-0911}{\includegraphics[scale=0.06]{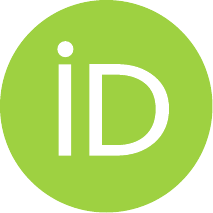}\hspace{1mm}David Klotz}\\ %\thanks{Use footnote for providing further
	Institute for Applied AI (IAAI)\\
	Media University\\
	Stuttgart, Germany \\
	\texttt{klotzd@hdm-stuttgart.de} \\
}
\else
\usepackage{authblk}

\newbox{\orcid}\sbox{\orcid}{\includegraphics[scale=0.06]{orcid.pdf}} 
\author[1]{%
	\href{https://orcid.org/0000-0000-0000-0000}{\usebox{\orcid}\hspace{1mm}David S.~Hippocampus\thanks{\texttt{hippo@cs.cranberry-lemon.edu}}}%
}
\author[1,2]{%
	\href{https://orcid.org/0000-0000-0000-0000}{\usebox{\orcid}\hspace{1mm}Elias D.~Striatum\thanks{\texttt{stariate@ee.mount-sheikh.edu}}}%
}
\affil[1]{Department of Computer Science, Cranberry-Lemon University, Pittsburgh, PA 15213}
\affil[2]{Department of Electrical Engineering, Mount-Sheikh University, Santa Narimana, Levand}
\fi

\renewcommand{\headeright}{}
\renewcommand{\shorttitle}{The Buy-or-Build Decision, Revisited}

\hypersetup{
	hidelinks,
	pdftitle={The Buy-or-Build Decision, Revisited: How Agentic AI Changes the Economics of Enterprise Software},
	pdfsubject={cs.CY},
	pdfauthor={David Klotz},
	pdfkeywords={Make-or-buy decision; Enterprise applications; Agentic coding systems; AI-augmented software development; IT strategy; AI governance},
}

\begin{document}
\maketitle

\begin{abstract}
	Advances in generative artificial intelligence, particularly agentic coding systems capable of autonomous software development, are disrupting the economics of the make-or-buy decision for enterprise applications. The ``SaaSocalypse'' narrative predicts that AI will render large segments of the Software-as-a-Service market obsolete by enabling firms to build software in-house at a fraction of historical cost. This paper adopts a conceptual research approach, combining transaction cost economics and the resource-based view with an assessment of current AI capabilities, to systematically re-evaluate the factors underlying the make-or-buy decision. It makes three contributions. First, it provides a factor-level analysis of how AI reshapes seven canonical decision determinants: cost, strategic differentiation, asset specificity, vendor lock-in, time-to-market, quality and compliance, and organizational capability. Second, it develops a typology of enterprise applications by their sensitivity to AI-induced shifts in make-or-buy economics. Third, it demonstrates that AI fundamentally transforms the governance properties of the Make option, shifting it from Williamson's pure hierarchy to a hybrid governance form that combines code ownership with external AI infrastructure dependency, with qualitatively different economics, capability requirements, and governance structures than pre-AI in-house development. The analysis finds that the SaaSocalypse thesis is overstated for most enterprise application categories; Make is most compelling for commodity utilities and differentiating custom applications in the AI era, while regulated and mission-critical systems remain predominantly in the buy domain.
\end{abstract}

% keywords can be removed

\section{Introduction} 

In early 2026, a wave of market anxiety swept through the enterprise software industry. Under the label ``SaaSocalypse,'' analysts and commentators warned that agentic artificial intelligence (AI) systems would enable firms to build previously purchased software in-house at a fraction of historical cost, rendering large segments of the Software-as-a-Service (SaaS) market obsolete \citep{TechCrunch2026SaaSocalypse, CNBC2026}. The narrative found dramatic expression in the capital markets: the S\&P Software \& Services Index declined by approximately 25\% in the first quarter of 2026, reflecting investor expectations that AI-driven insourcing would erode SaaS vendors' recurring revenue streams \citep{Bloomberg2026}.

The underlying thesis is seductive in its simplicity. Agentic coding systems---AI agents capable of end-to-end code generation, testing, debugging, and deployment---have advanced rapidly \citep{Anthropic2025, OpenAICodex2025, WangEtAl2025, CrewAI2024}. Where software development once required large teams working over months, small teams augmented by AI agents can now deliver functional applications in days or weeks. If the primary barrier to building enterprise software in-house was cost and time, and AI dismantles both barriers, then the rationale for purchasing standardized SaaS solutions appears to weaken fundamentally. Yet, this reasoning collides with well-established knowledge in information systems (IS) research. A widely cited heuristic of software lifecycle costs holds that initial development accounts for a minority of total cost of ownership (TCO), while operations, maintenance, enhancement, and eventual retirement consume the majority, commonly estimated at 60--80\% of lifecycle costs, \citep{LientzSwanson1980, Boehm1981, Glass2002}. Even if AI significantly reduces the development share, the operational majority of lifecycle costs remains and may in fact grow more complex when AI-generated codebases require governance, quality assurance, and ongoing adaptation. The SaaSocalypse thesis, in other words, may address the smaller part of the cost equation while underestimating the larger one. A further complication arises from the competitive dynamics of the supply side. The same generative AI capabilities that empower firms to build software in-house are equally available to SaaS vendors. Established providers are embedding AI natively into their platforms, accelerating release cycles, deepening feature sets, and, critically, expanding switching costs through AI-powered personalization and workflow integration \citep{CusumanoEtAl2019, ParkerEtAl2016}. The make-or-buy calculus is thus not a one-sided shift toward ``make'' but a simultaneous transformation of both options, with the net outcome depending on application-specific and firm-specific factors.

These dynamics call for a systematic re-examination of the Make-or-Buy (MoB) decision for enterprise applications in light of generative AI. The MoB decision has been studied extensively in IS through the lenses of transaction cost economics \citep[TCE;][]{Williamson1981, Williamson1985}, the resource-based view \citep[RBV;][]{Barney1991, MataEtAl1995}, and practical frameworks such as total cost of ownership analysis and IT portfolio management \citep{LacityHirschheim1993, KernWillcocks2001}. However, these frameworks predate the emergence of agentic AI systems and do not account for the profound shifts in cost structure, capability requirements, and risk profiles that such systems introduce.

This paper addresses the following research question: \emph{How do advances in generative AI, particularly agentic coding systems, alter the factors underlying the Make-or-Buy decision for enterprise applications, and what revised decision logic follows for IS strategy?}

The paper makes three contributions. First, it provides a systematic, factor-level analysis of how AI reshapes the canonical determinants of the MoB decision: cost, strategic differentiation, asset specificity, vendor lock-in, time-to-market, quality and compliance, and organizational capability. Second, it develops a typology of enterprise applications by their sensitivity to AI-induced MoB shifts, distinguishing commodity utilities, differentiating custom applications, regulated standard applications, and mission-critical systems of record. Third, it argues that AI fundamentally transforms the governance properties of the Make option, shifting it from Williamson's pure hierarchy to a hybrid governance form that combines code ownership with external AI infrastructure dependency, and proposes a revised binary decision framework that accounts for this transformation and its implications for IS strategy. To address these contributions, the study adopts a conceptual research approach \citep{Jaakkola2020}, combining deductive analysis grounded in established IS theories (TCE, RBV) with an assessment of emergent AI capabilities to derive a revised decision framework. Specifically, each canonical MoB decision factor is systematically re-evaluated against the capabilities and limitations of current agentic AI systems. This approach is appropriate given the nascent state of AI-augmented development practices, which precludes large-scale empirical investigation at this stage.

The remainder of the paper is organized as follows. Section~2 reviews the theoretical background on MoB decision frameworks and the current state of generative AI in software development. Section~3 analyzes how AI transforms each of the seven key MoB decision factors. Section~4 presents the revised decision framework, the application typology, and the updated decision logic. Section~5 discusses implications, practical challenges, and limitations. Section~6 concludes.

\section{Background}

This section establishes the theoretical and technical foundations for the subsequent analysis. Section~2.1 reviews the canonical frameworks that have guided MoB decisions in IS, synthesizing them into a set of decision factors. Section~2.2 surveys the state of generative AI in software development, with particular attention to agentic coding systems and their capability boundaries.

\subsection{Make-or-Buy Decision Frameworks in Information Systems}

\subsubsection{Transaction Cost Economics}

The theoretical foundation for MoB decisions in IS derives primarily from transaction cost economics (TCE). \citet{Coase1937} posed the foundational question of why firms exist at all if markets can coordinate economic activity through the price mechanism, answering that firms arise when the costs of organizing transactions internally fall below the costs of conducting them through markets. \citet{Williamson1981, Williamson1985} operationalized this insight by identifying three key attributes of transactions that determine efficient governance: \emph{asset specificity}, the degree to which investments are specialized to a particular transaction; \emph{uncertainty}, the difficulty of specifying and monitoring contractual terms; and \emph{transaction frequency}, how often the transaction recurs.

In Williamson's framework, higher asset specificity favors internal governance (``hierarchy'') because specialized investments create bilateral dependency and expose the investing party to hold-up risks. Conversely, transactions involving standardized assets are efficiently governed through markets. \citet{Williamson1991} extended this analysis by formalizing three discrete structural alternatives: \emph{market} governance (arm's-length contracts, high-powered incentives), \emph{hierarchy} (internal organization, administrative controls), and \emph{hybrid} forms that combine elements of both, such as long-term partnerships, joint ventures, or relational contracts that share ownership rights and blend market incentives with hierarchical coordination. More broadly, hybrid governance arises whenever a single value-creating activity combines hierarchical control over some elements with market-based procurement of others (a configuration that, as Section~4 argues, characterizes AI-augmented development arrangements).

Applied to IS, TCE has been the dominant lens for analyzing outsourcing decisions \citep{LacityHirschheim1993, LacityEtAl2009}. The central prediction that highly firm-specific applications should be developed in-house while generic requirements favor external procurement has been broadly supported empirically, though moderated by factors such as measurement difficulty, technological uncertainty, and vendor market maturity \citep{KernWillcocks2001}.

\subsubsection{Resource-Based View}

The resource-based view (RBV) provides a complementary perspective grounded in strategic management theory. \citet{Barney1991} argued that sustained competitive advantage derives from resources that are valuable, rare, inimitable, and non-substitutable (the VRIN criteria). Applied to IS, the RBV shifts the make-or-buy question from transaction cost minimization to strategic value maximization: if an application constitutes or enables a core competency, retaining it in-house preserves the firm's ability to appropriate its strategic value. \citet{MataEtAl1995} examined which IT resources satisfy the VRIN criteria. They concluded that technical IT resources (hardware, proprietary software) are generally imitable and thus unlikely to confer sustainable advantage. Instead, \emph{IT management capability} (the organizational ability to conceive, develop, and exploit IT applications) emerged as the most promising source of IT-based competitive advantage, precisely because it is socially complex and path-dependent. \citet{WadeHulland2004} reinforced this finding in a comprehensive review, identifying IT-related resources such as IS planning sophistication, development capability, and IS--business partnerships as key competitive assets.

The RBV thus provides the theoretical basis for a ``make'' recommendation that goes beyond TCE's cost calculus: even when external procurement is cheaper, in-house development may be strategically warranted if the application is integral to a capability that differentiates the firm. This argument gains new significance in the AI era, where the nature of the differentiating capability itself may shift from application development skill to AI orchestration capability, a possibility examined in Section~4.

\subsubsection{IS-Specific Frameworks and Decision Factor Synthesis}

Beyond TCE and RBV, several practitioner-oriented frameworks inform the make-or-buy decision in IS. The concept of \emph{total cost of ownership} (TCO) provides a lifecycle perspective on application economics. The widely cited heuristic holds that operations, maintenance, enhancement, and retirement consume the majority of an application's total lifecycle cost (estimates range from 40\% to 80\%, with 60--80\% being the most commonly cited range) \citep{LientzSwanson1980, Boehm1981, Glass2002}. This cost asymmetry has historically strengthened the case for ``buy,'' since enterprise software vendors amortize development, operational, and compliance costs across their customer base, regardless of whether they deliver solutions as SaaS subscriptions, private cloud deployments, or traditional on-premise installations. Yet lifecycle economics alone do not determine governance mode; the strategic character of an application is equally consequential.

The provocative claim that IT had become a commodity input, necessary but insufficient for competitive advantage, sparked considerable debate \citep{Carr2003}, yet it usefully highlights the distinction between \emph{commodity IT} (standardized, widely available) and \emph{strategic IT} (firm-specific, competitively relevant). An early response to this heterogeneity was to treat IT investments as a portfolio, balancing risk and strategic contribution \citep{McFarlan1981}. Empirical support for deliberate portfolio allocation followed: firms distributing IT investment across infrastructure, transactional, informational, and strategic categories were found to achieve superior returns \citep{Weill2006}. Portfolio thinking thus reframes the MoB question from a cost calculation to a strategic positioning exercise, a shift that the pace-layering model \citep{Gartner2012} subsequently operationalized at the application level. The model provides a complementary classification that has become widely adopted in enterprise IT strategy and governance, distinguishing three application tiers by their rate of change, each with distinct implications for the MoB decision:

\begin{enumerate}
	\item \emph{Systems of record}---such as ERP core modules, core banking platforms, and HR administration systems---are characterized by stability, high compliance requirements, and standardized processes. Their long lifecycles and stringent auditability needs have historically made them strong candidates for ``buy,'' as vendors can amortize compliance and maintenance costs across large customer bases.
	\item \emph{Systems of differentiation}---including CRM extensions, supply chain optimization tools, and proprietary analytics platforms---support firm-specific business processes and operate on shorter change cycles. These applications occupy the boundary zone between ``make'' and ``buy'': their competitive relevance argues for in-house control, while their complexity often favors vendor solutions.
	\item \emph{Systems of innovation}---such as prototypes, experimental customer interaction channels, and novel AI-powered features---are characterized by short lifespans, high rates of change, and tolerance for failure. These are natural ``make'' candidates, as the speed and flexibility of in-house development outweigh the benefits of standardized procurement.
\end{enumerate}

The pace-layering perspective reinforces a key insight for the present analysis: the MoB decision is not a uniform, portfolio-wide choice but must be differentiated by application layer. This layer-specific logic informs the more granular application typology developed in Section~4.2, which extends the three-tier model by incorporating compliance exposure and domain specificity as additional dimensions.

These conceptual frameworks converge on a set of theoretical predictions that the IS outsourcing literature has subjected to systematic empirical scrutiny. In the most comprehensive meta-analysis of this literature, \citet{LacityEtAl2009} synthesized findings from 191 articles spanning two decades of IS outsourcing research. Their analysis identified asset specificity, strategic importance, and environmental uncertainty as the most robust and consistently significant predictors of governance mode choice. Beyond these core predictors, the study found that contractual governance mechanisms (e.g., contract detail, flexibility provisions) and relational governance (trust, communication quality) both independently influence outsourcing success, a finding that resonates with the hybrid governance perspective developed in Section~4. The meta-analysis also confirmed that cost reduction alone is an insufficient basis for outsourcing decisions; firms that outsource primarily for cost savings report lower satisfaction than those guided by strategic considerations.

Synthesizing across these theoretical and practical perspectives, seven factors emerge as the canonical determinants of the MoB decision for enterprise applications:

\begin{enumerate}
    \item \textbf{Cost structure (TCO):} The total lifecycle cost comparison between internal development and external procurement, comprising development expenditure (CAPEX) and ongoing operational cost (OPEX). Rooted in TCE's efficiency logic and the lifecycle cost heuristic.
    \item \textbf{Strategic differentiation potential:} The degree to which an application contributes to competitive advantage. Derived from the RBV and the strategic-vs-commodity IT distinction.
    \item \textbf{Asset specificity and customization requirements:} The extent to which the application must be tailored to firm-specific processes, data, or domain logic. The central TCE variable.
    \item \textbf{Vendor dependency and lock-in risk:} The switching costs and strategic exposure associated with reliance on an external provider's platform, data formats, and APIs. Informed by TCE (bilateral dependency) and the lock-in literature \citep{Shapiro1999, OparaEtAl2017}.
    \item \textbf{Time-to-market and organizational agility:} The speed at which the application can be deployed and subsequently adapted. Linked to TCE's uncertainty dimension---in dynamic environments, the ability to respond rapidly reduces exposure to environmental change---and to the agility requirements emphasized in the IS outsourcing literature.
    \item \textbf{Quality, reliability, and compliance requirements:} The standards for correctness, availability, security, and regulatory conformity that the application must meet. Rooted in TCE's concern with measurement difficulty and monitoring costs; in regulated industries, the burden of demonstrating compliance creates governance costs that favor certified external solutions.
    \item \textbf{Organizational capability:} The firm's internal capacity to develop, operate, and govern the application. Encompasses technical skills, management practices, and institutional knowledge. Grounded in the RBV's emphasis on management capability \citep{MataEtAl1995, WadeHulland2004}.
\end{enumerate}

These seven factors provide the analytical structure for the assessment in Section~3, where each is re-evaluated in light of generative AI capabilities.

\subsection{Generative AI and Agentic Coding Systems}

\subsubsection{From Language Models to Code Generation}

The technical foundation for AI-augmented software development lies in large language models (LLMs), large-scale neural models trained on extensive corpora of natural language and source code. The application of LLMs to software engineering has progressed through several stages. Early code-generation models demonstrated the ability to produce syntactically correct function implementations from natural language descriptions \citep{ChenEtAl2021}. Subsequent integration into developer environments as ``copilot'' assistants providing real-time code suggestions, completions, and documentation, yielding measurable productivity gains: \citet{PengEtAl2023} reported that developers using GitHub Copilot completed tasks approximately 55\% faster than a control group, with the effect most pronounced for well-specified, self-contained coding tasks.

Comprehensive surveys document the breadth of LLM applications in software engineering, spanning code generation, testing, debugging, code review, documentation, and program repair \citep{HouEtAl2024, FanEtAl2025, YangEtAl2025Survey}. The consistent finding across this literature is that LLMs are most effective for tasks that are well-specified, modular, and similar to patterns represented in training data, while performance degrades for tasks requiring deep domain reasoning, cross-system integration, or handling of ambiguous requirements.

\subsubsection{Agentic Coding Systems: Definition and State of the Art}

A qualitative shift beyond copilot-style assistance has emerged with \emph{agentic coding systems} (AI systems capable of autonomous, end-to-end software development tasks with minimal human intervention). It is useful to distinguish an ``AI agent'' (a single autonomous system pursuing goals in an environment) from the broader paradigm of ``agentic AI,'' in which LLM-based systems autonomously plan, execute, and iterate on complex tasks \citep{SapkotaEtAl2026}; agentic coding systems are a specific instantiation of agentic AI in the software engineering domain. Where copilot tools respond to individual prompts within a developer's workflow, agentic systems operate in closed loops: they interpret a high-level task specification, plan an implementation strategy, generate code, execute tests, diagnose failures, and iterate toward a solution \citep{YangEtAl2024}.

Representative systems illustrate the current state of the art. SWE-agent \citep{YangEtAl2024} demonstrated autonomous resolution of real-world GitHub issues from the SWE-bench benchmark \citep{JimenezEtAl2024}, which evaluates agents against actual bug reports and feature requests drawn from popular open-source repositories. A growing ecosystem of systems extends these capabilities to production software development. Integrated coding agents (including Claude Code \citep{Anthropic2025, ChatlatanagulchaiEtAl2025}, OpenAI Codex \citep{OpenAICodex2025}, GitHub Copilot Workspace \citep{GitHubCopilot2024}, Gemini Code Assist \citep{GeminiCodeAssist2025}, Devin \citep{Cognition2024}, Cursor, OpenHands \citep{WangEtAl2025}, and Aider \citep{Aider2024}) offer end-to-end workflows encompassing code generation, refactoring, test authoring, debugging, and deployment configuration. Complementing these are multi-agent orchestration frameworks such as AutoGen \citep{WuEtAl2024}, CrewAI \citep{CrewAI2024}, and LangChain/LangGraph \citep{LangChain2023}, which enable firms to compose custom agentic workflows by coordinating multiple specialized agents within a single development pipeline. The pace of real-world adoption corroborates the technical narrative: an empirical analysis of over 129,000 public GitHub projects found that between 16\% and 23\% had already integrated agentic coding tools within months of their general availability, spanning projects at all maturity levels, established organizations, and diverse programming languages and domains \citep{RobbesEtAl2026}.

The key capabilities of current agentic systems include: (1)~greenfield application development from natural language specifications; (2)~automated test generation and execution; (3)~bug diagnosis and patch generation; (4)~code refactoring and modernization; and (5)~emerging closed-loop operation in which agents monitor deployed applications, detect faults, and generate corrective patches autonomously.

\subsubsection{The Automation Frontier: Capabilities and Limitations}

Despite rapid advances, agentic coding systems exhibit systematic capability boundaries that are directly relevant to the MoB decision. Current systems perform well on tasks that are \emph{well-specified} (clear requirements, unambiguous acceptance criteria), \emph{modular} (limited scope, weak coupling to other components), and \emph{pattern-rich} (similar to structures well-represented in training data). This profile aligns with greenfield implementations of standard functionality, CRUD applications, reporting tools, API integrations, test automation, and routine maintenance tasks \citep{RahmanEtAl2025, HorikawaEtAl2025}. Conversely, significant limitations persist for tasks involving: (1)~complex legacy system migration, where undocumented interdependencies and implicit business rules resist formal specification; (2)~tacit domain logic that resides in organizational practices rather than code or documentation \citep{ZhouEtAl2025}; (3)~cross-organizational processes requiring negotiation of competing stakeholder requirements; (4)~high-stakes correctness requirements where subtle errors carry disproportionate consequences, such as financial calculations, safety-critical systems, or regulatory reporting; and (5)~multi-system integration requiring consistent data models, API contracts, and transactional integrity across enterprise architectures such as ERP, CRM, and data warehouse systems \citep{RahmanEtAl2025, MiserendinoEtAl2025}.

These capability boundaries have a direct bearing on the lifecycle cost heuristic introduced in Section~2.1. Agentic systems primarily compress the development cost share, and they begin to erode the dominant operational and maintenance cost share through automated maintenance, monitoring, and bug-fixing. However, they do not eliminate the operational share: governance, quality assurance, security oversight, and adaptation to evolving requirements remain human-intensive activities \citep{SunEtAl2025}. The net effect is a shift in the cost distribution rather than a collapse of lifecycle costs, a shift that nonetheless has significant implications for the MoB decision factors synthesized in Section~2.1, as analyzed in detail in Section~3.

\section{Analysis: How AI Transforms the MoB Decision Factors}

This section re-evaluates each of the seven canonical decision factors identified in Section~2.1 in light of the AI capabilities and limitations established in Section~2.2. For each factor, the analysis considers both the effect on in-house development and the simultaneous transformation of external procurement, since the net shift depends on the relative magnitude of change on both sides.

\subsection{Cost Structure (TCO)}

\paragraph{Effect on Make.} Agentic coding systems substantially reduce the development cost share (CAPEX) of in-house applications. Emerging evidence suggests that tasks which previously required dedicated teams working over months (such as greenfield implementations, standard integrations, and reporting tools) can be completed substantially faster by small teams augmented by AI agents \citep{PengEtAl2023, JimenezEtAl2024}. The operational cost share (OPEX) is also partially compressed as AI agents automate routine maintenance, bug-fixing, monitoring, and test regression. However, new cost categories emerge: AI infrastructure expenditure (model API costs, compute), governance overhead for AI-generated artifacts, quality assurance processes tailored to AI-produced code, and integration costs for connecting AI-built applications to existing enterprise system landscapes. The net CAPEX reduction is substantial; the net OPEX reduction is moderate and partially offset by these new cost categories.

\paragraph{Effect on Buy.} Enterprise software vendors across all delivery models leverage the same AI capabilities to reduce their own development and operational costs. In principle, these savings could be passed to customers through lower prices. In practice, vendors are more likely to reinvest savings into AI-native features, deeper functionality, and tighter platform integration, thereby increasing perceived value rather than reducing price. The vendor's amortization advantage (spreading development costs across a large customer base) persists regardless of whether software is delivered as SaaS, private cloud, or on-premise, though its relative importance diminishes as AI reduces the absolute cost of building alternatives.

\paragraph{Net effect.} The cost asymmetry that historically favored ``buy'' weakens considerably for lower-complexity applications where AI's development productivity gains are largest. For applications requiring high operational reliability, complex domain knowledge, or extensive compliance infrastructure, the vendor's scale advantages in operational cost amortization remain significant. The cost factor thus becomes more \emph{application-dependent} than in the pre-AI era: a strong shift toward ``make'' for commodity and simple custom applications, a modest shift for moderately complex systems, and minimal change for mission-critical systems of record.

\subsection{Strategic Differentiation and Core Competency (RBV)}

\paragraph{Effect on Make.} As build costs fall, the set of applications that can economically serve as sources of competitive advantage expands. Applications that were previously too expensive to justify custom development (customer-facing experiences, proprietary analytics, process-specific automation) now become viable ``make'' candidates. Moreover, the same cost reduction applies to customizations and extensions of standard software packages such as SAP or Salesforce: enhancements that were previously cost-prohibitive become viable when AI dramatically reduces implementation effort. The locus of the differentiating asset shifts from the application itself toward the firm's capability to orchestrate AI-augmented development. Building this capability becomes a strategic investment in its own right.

\paragraph{Effect on Buy.} Established vendors respond by embedding AI to offer deeper customization, AI-powered analytics, and more adaptive workflows within their products. This applies across delivery models: SaaS platforms add personalization layers, while on-premise and private cloud vendors such as SAP and Oracle are embedding AI capabilities natively into their core systems. These investments partially close the differentiation gap, enabling firms to achieve within a ``buy'' framework a degree of customization that was previously available only through in-house development. AI-powered platform ecosystems may themselves become sources of network-effect-based differentiation \citep{CusumanoEtAl2019, ParkerEtAl2016}.

\paragraph{Net effect.} The differentiation argument shifts in favor of ``make'' for applications where firm-specific logic, data, or processes are central to competitive advantage. However, a \emph{rarity paradox} emerges: if AI enables every firm to build custom solutions at low cost, the applications themselves may cease to satisfy the RBV's rarity criterion for sustained competitive advantage \citep{Barney1991}. The differentiating resource then becomes not the application but the organizational AI development capability, which, being socially complex and path-dependent, may satisfy the VRIN criteria more readily than any individual application \citep{MataEtAl1995, WadeHulland2004}. This capability is path-dependent because it encompasses organization-specific experience in managing AI outputs, proprietary prompt libraries and validation workflows, and established quality assurance processes for AI-generated code, assets that are built incrementally and resist straightforward imitation.

\subsection{Asset Specificity and Customization Requirements (TCE)}

\paragraph{Effect on Make.} Asset specificity has been the primary TCE argument for ``make'': when requirements are highly firm-specific, external procurement entails hold-up risks and costly adaptation of standardized solutions. AI reduces the cost of producing firm-specific applications, effectively lowering the \emph{specificity threshold} at which in-house development becomes economically rational. AI excels particularly at well-specified customization where requirements can be articulated clearly, even if they are highly specific.

\paragraph{Effect on Buy.} Enterprise software vendors invest heavily in configurability, low-code extension mechanisms, and AI-powered customization layers to reduce the ``specificity gap'' between standardized solutions and firm-specific needs. However, architectural constraints on deep customization persist in all cases. Multi-tenant SaaS platforms cannot accommodate arbitrary process redesigns or proprietary data models without compromising scalability and maintainability. On-premise products from vendors such as SAP or Oracle permit deeper customization but at significant implementation and maintenance cost, as highly tailored configurations introduce long-term rigidities of their own.

\paragraph{Net effect.} The net shift favors ``make'' for applications with moderate to high specificity, where AI can efficiently translate well-articulated requirements into custom implementations. The residual challenge lies in requirements that are \emph{tacit}, \emph{evolving}, and \emph{contradictory}, characteristics that resist formal specification and thus limit AI's ability to automate custom development. For such requirements, neither ``buy'' (insufficient customization) nor AI-assisted in-house development (insufficient specification) provides an ideal solution; deep domain expertise and careful human-driven design retain their role.

\subsection{Vendor Dependency and Lock-in Risk}

\paragraph{Effect on Make.} In-house development, whether traditional or AI-augmented, yields code ownership, which provides structural protection against application-level lock-in. The firm retains control over its data, APIs, and business logic. However, AI-assisted in-house development introduces a new form of dependency: reliance on AI infrastructure providers for model APIs, cloud compute, and development tooling. This ``AI infrastructure lock-in'' is qualitatively different from SaaS application lock-in: infrastructure dependencies are more fungible (multiple providers offer comparable capabilities), the switching costs are lower (code remains portable), and the dependency operates at a lower level of abstraction.

\paragraph{Effect on Buy.} Vendor lock-in mechanisms remain in force across all delivery models and arguably deepen as vendors embed AI features that create additional switching costs. For SaaS platforms, these mechanisms include proprietary data formats, API dependencies, and workflow integration. For on-premise systems from vendors such as SAP or Oracle, lock-in is often more deeply entrenched, arising from proprietary customization frameworks, large sunk costs in implementation and training, and organizational processes built around vendor-specific workflows. AI-powered personalization and vendor-specific AI assistants add a further layer of switching cost in both contexts \citep{Shapiro1999, OparaEtAl2017}.

\paragraph{Net effect.} Make in the AI era trades application-level lock-in for infrastructure-level lock-in. The latter is generally less binding: infrastructure is more commoditized, switching costs are lower, and the firm retains its most strategically valuable asset: the codebase and its embedded business logic. The lock-in argument thus shifts moderately in favor of ``make,'' though firms must actively manage their AI infrastructure dependencies to realize this advantage.

\subsection{Time-to-Market and Organizational Agility}

\paragraph{Effect on Make.} Agentic coding systems compress development timelines dramatically for well-scoped applications. For greenfield projects with clear requirements, AI-augmented teams can achieve deployment speeds approaching those of SaaS procurement, eliminating one of the traditional ``buy'' arguments. Ongoing adaptation is similarly accelerated: feature additions, modifications, and refactoring that previously required sprint cycles can be accomplished in a fraction of the time previously required. However, integration with existing enterprise systems (ERP, CRM, identity management, data warehouses) can significantly extend deployment timelines even when the core application is built rapidly, as cross-system integration requires negotiating API contracts, data model alignment, and organizational coordination that AI cannot fully automate (cf.\ the integration limitation identified in Section~2.2).

\paragraph{Effect on Buy.} Established vendors leverage AI to accelerate their own development and release cycles, enabling faster feature delivery and more rapid response to customer demand. For mature products in established categories, the cumulative advantage of years of development, customer feedback, and feature refinement represents a time-to-value lead that AI-augmented greenfield development cannot instantly match. This advantage is particularly pronounced for large on-premise platforms with broad installed bases, where accumulated domain logic, pre-built integrations, and ecosystem partnerships represent substantial embedded value. Notably, however, system-of-record applications benefit little from faster vendor innovation cycles: these platforms are typically mature, with limited demand for new business-facing features. Enterprise customers in this segment tend to favor stability over agility, organizing operations around infrequent, carefully governed release windows with only a handful of major changes per year. For such applications, accelerated vendor delivery is largely irrelevant, weakening one of the traditional arguments for ``buy.''

\paragraph{Net effect.} The time-to-market advantage of ``buy'' diminishes for applications in early-stage or fast-moving categories, where AI-augmented development can deliver functional solutions before a suitable vendor offering matures. For established categories with mature vendor products, the accumulated feature depth and ecosystem integration of commercial platforms continue to provide a time-to-value advantage. The net effect is category-dependent: strongest shift toward ``make'' for novel application types, weakest for commoditized categories with mature vendor ecosystems.

\subsection{Quality, Reliability, and Compliance Requirements}

\paragraph{Effect on Make.} AI-generated code introduces quality risks that differ qualitatively from those of human-authored code: subtle logic errors that pass superficial review, security vulnerabilities arising from pattern-matching without semantic understanding, and hallucinated dependencies on non-existent libraries or APIs \citep{HouEtAl2024}. A ``last mile'' problem characterizes current agentic systems: they resolve the majority of well-specified requirements efficiently, but edge cases, failure modes, and domain-specific correctness requirements demand expert human oversight. On the SWE-bench Verified benchmark, which evaluates agents against real-world GitHub issues, leading systems now resolve approximately 80\% of tasks \citep{JimenezEtAl2024, SWEBenchLeaderboard2026}, a dramatic improvement from the 13.86\% achieved by early agentic systems in 2024 \citep{Cognition2024}. Yet empirical studies find that a majority of AI-generated code contains security vulnerabilities: \citet{TihanyiEtAl2024} report that over 62\% of LLM-generated programs exhibit exploitable weaknesses. The remaining unresolved benchmark tasks and the persistent security risks disproportionately involve cross-file reasoning, complex domain logic, and subtle correctness requirements, precisely the characteristics that define mission-critical enterprise applications. Compliance documentation for AI-generated artifacts (audit trails, provenance records, and conformity assessments) remains an immature practice.

\paragraph{Effect on Buy.} Established enterprise software vendors offer certified, auditable solutions with track records of regulatory compliance. This applies equally to SaaS providers and to vendors delivering on-premise or private cloud deployments, such as SAP and Oracle and their equivalents in vertical markets. Compliance documentation, data lineage, liability attribution, and security certification are core competencies of these providers, particularly in regulated industries, and represent years of accumulated investment that is difficult to replicate with AI-augmented in-house development.

\paragraph{Net effect.} The quality and compliance factor strongly favors ``buy'' in regulated contexts. Sectoral regulations, e.g., DORA \citep{DORA2022} for financial services, NIS2 \citep{NIS2022} for critical infrastructure, MDR \citep{MDR2017} for medical devices, impose certification and documentation requirements that established vendors have invested years in satisfying and that AI-augmented in-house development cannot replicate quickly. Beyond regulatory certification, AI-augmented make introduces a distinct liability dimension. When custom-developed software fails in production, the firm bears the full consequences: service disruptions, lost revenue, reputational damage, and customer attrition. Where a software failure violates legal or regulatory requirements, the organization and its responsible officers may also face direct legal liability. AI coding tools and their vendors explicitly disclaim responsibility for the correctness or lawfulness of generated code; quality assurance remains entirely with the commissioning firm. While AI can assist with legal analysis and compliance checking in the same way it assists with coding, this does not transfer liability: the firm retains accountability for the lawfulness of its custom applications. For low-compliance, non-critical applications the balance is more even, as the production risk is contained and AI-generated code quality is often adequate.

\subsection{Organizational Capability Requirements}

\paragraph{Effect on Make.} AI-assisted in-house development requires a capability profile that differs qualitatively from both pre-AI in-house development and external software procurement. Rather than a large development workforce with conventional programming expertise, it demands skills in prompt engineering, agent orchestration, AI output validation, and governance of AI-generated artifacts. This capability gap represents a structural barrier: not all firms can acquire or develop these skills in the near term, and the talent market for AI orchestration expertise is highly competitive.

\paragraph{Effect on Buy.} Procuring from established vendors requires minimal internal development capability, which has been a key advantage for firms without strong IT departments. This holds across delivery models: SaaS subscriptions, private cloud deployments, and traditional on-premise installations all absorb technical complexity on the vendor side. However, reliance on ``buy'' forecloses the development of internal AI capabilities, potentially creating a long-term strategic dependency: firms that exclusively buy may lack the skills to evaluate, govern, or eventually transition to AI-augmented in-house alternatives.

\paragraph{Net effect.} Organizational capability acts as a \emph{threshold variable}: below a minimum level of AI development capability, Make via AI tools is not viable regardless of how favorable the other factors may be. Above this threshold, the capability investment compounds over time, as AI orchestration skill, being socially complex and path-dependent (as argued in Section~3.2), satisfies the RBV's criteria for a sustainable competitive resource \citep{MataEtAl1995, WadeHulland2004}. The capability factor thus introduces a dynamic dimension to the MoB decision: it favors ``buy'' in the short term for capability-constrained firms, but favors ``make'' in the long term for firms that invest in AI development capability.

\section{A Revised Decision Framework for the AI Era}

\subsection{The Transformation of Make: From Hierarchy to Hybrid}

The preceding analysis reveals that AI does not simply shift the balance between ``buy'' and ``make''; it fundamentally transforms the governance properties of the Make option itself. Where pre-AI in-house development embodied Williamson's \emph{hierarchy} governance form, characterized by internal code ownership, direct control over application logic, and a large, long-term development workforce, AI-era Make acquires market-like properties that shift it along the governance spectrum toward Williamson's \emph{hybrid} form \citep{Williamson1991}.

\paragraph{Theoretical grounding in TCE.} The hierarchy-like properties of Make are retained in the AI era: the firm continues to own its codebase, control its application logic, and govern its development process internally. What changes is the production infrastructure. AI-era Make depends on external AI providers for model APIs, cloud compute, and development tooling, all priced on variable, pay-per-use terms rather than the fixed employment and infrastructure commitments of traditional development. Enterprises can partially mitigate this external dependency by deploying open-source AI models on internal infrastructure, whether on-premises or on private cloud, thereby retaining control over the production environment. This approach comes with trade-offs, however: open-source models currently lag behind the frontier capabilities of leading commercial providers, and internally hosted deployments tend to receive capability improvements later, as the pace of AI innovation is driven predominantly by commercial vendors. In either case, a dependency to the AI model vendor remains. This combination of hierarchical control over the application artifact with market-like flexibility in the production infrastructure constitutes, in Williamson's terms, a hybrid governance arrangement \citep{Williamson1991}: a structural shift in the nature of Make, not merely a cost variation.

This shift has analytical significance beyond the choice of production inputs. AI infrastructure dependency exhibits structural properties that set AI-era Make apart from conventional tool usage. First, when an external AI provider is used, model APIs represent \emph{variable production inputs} (pay-per-token pricing) rather than fixed infrastructure costs, meaning that the marginal cost of each development activity depends directly on an external provider's pricing, a market-like property embedded within a hierarchical activity. Second, external AI providers gain visibility not only into application data but into the firm's \emph{business logic, requirements, and source code}, since the AI system is an active participant in the production process rather than a passive infrastructure layer, creating an information asymmetry qualitatively deeper than conventional cloud hosting. Both of these properties can be avoided by deploying AI models on internal infrastructure, though as noted above this typically comes at the cost of sub-standard model outputs. Third, AI production processes are \emph{non-deterministic}: identical inputs may yield different outputs, necessitating a dedicated governance layer (validation, quality assurance, human oversight) that has no analogue in deterministic development tooling. Fourth, the \emph{capability envelope} of AI systems shifts rapidly, often quarterly, making the governance arrangement inherently dynamic in ways that stable infrastructure services are not.

\paragraph{Theoretical grounding in RBV.} From a resource-based perspective, AI transforms the capability bundle that the Make option requires \citep{MataEtAl1995, WadeHulland2004}: from a large software engineering workforce with conventional programming expertise to AI orchestration, prompt engineering, agent governance, and quality assurance for AI-generated artifacts. AI-era Make is therefore not a cost-optimized variant of the same resource configuration but a qualitatively different capability investment, one whose strategic value derives from the accumulation of AI orchestration expertise rather than domain-specific development skill. As argued in Section~3.2, this capability satisfies the RBV's VRIN criteria more readily than any individual application, since it is socially complex, path-dependent, and built incrementally through organizational experience. Table~\ref{tab:three-options} summarizes the governance properties of the Buy and Make options across the pre-AI and AI eras, illustrating the transformation of Make's governance structure.

\begin{table}[ht]
\centering
\caption{The transformation of the Make option in the AI era: comparison with Buy and pre-AI Make.}
\label{tab:three-options}
\small
\begin{tabular}{p{3.8cm}p{3.2cm}p{3.2cm}p{3.8cm}}
\hline
\textbf{Property} & \textbf{Buy} & \textbf{Make (Pre-AI Era)} & \textbf{Make (AI Era)} \\
\hline
Code ownership & No & Yes & Yes \\
Customization freedom & Limited & Full & High (within AI capability) \\
Upfront development cost & Low & High & Medium-low \\
Operational cost & Subscription-based & High (team + infra) & Low-medium (agents + infra) \\
Time-to-market & Fast & Slow & Fast-medium \\
AI infrastructure dependency & Indirect & None & Direct \\
Required internal capability & Low & High (traditional dev) & New (AI orchestration) \\
Compliance auditability & High (certified) & High (controlled) & Medium (emerging practices) \\
TCE governance mode & Market & Hierarchy & Hybrid \\
\hline
\end{tabular}
\end{table}

\subsection{A Typology of Enterprise Applications by MoB-AI Sensitivity}

The factor-level analysis in Section~3 demonstrates that the magnitude of AI-induced change varies substantially across application types. To operationalize this insight, we propose a typology that classifies enterprise applications along three analytically grounded dimensions:

\begin{enumerate}
    \item \textbf{Application complexity} (derived from RBV): the scope of functional requirements, integration depth, and criticality of correctness. Higher complexity implies greater resource commitment and higher stakes for the firm's competitive position, increasing the strategic weight of the differentiation and capability factors.
    \item \textbf{Domain specificity} (derived from TCE): the degree to which requirements are firm-specific versus generic across the industry. This dimension maps directly onto Williamson's asset specificity \citep{Williamson1985}, the primary TCE determinant of governance choice.
    \item \textbf{Compliance exposure} (derived from TCE): the extent of regulatory, legal, or auditability obligations. Compliance exposure operationalizes a specific, practically relevant manifestation of environmental uncertainty in regulated industries: the uncertainty not about market conditions but about regulatory interpretation, audit outcomes, and liability attribution.
\end{enumerate}

These dimensions yield four application categories with distinct MoB-AI sensitivity profiles. While three dimensions theoretically permit a larger set of combinations, the four categories identified below represent the empirically most prevalent application profiles in enterprise IT portfolios and capture the principal variation in AI-induced MoB sensitivity:

\paragraph{Category 1: Commodity utilities} (low complexity, low specificity, low compliance). Examples include internal tooling, simple reporting dashboards, and basic workflow automation. AI-induced MoB shift: \emph{high}. These applications lie squarely within the capability envelope of current agentic systems; their low specificity and minimal compliance requirements remove the two strongest arguments for ``buy.'' Make becomes the default recommendation; the case for Buy weakens most sharply in this category.

\paragraph{Category 2: Differentiating custom applications} (low-medium complexity, high specificity, low-medium compliance). Examples include customer-facing features, proprietary process automation, and competitive analytics. AI-induced MoB shift: \emph{high}. The combination of high specificity (favoring ``make'' under TCE) and low-medium compliance (removing regulatory barriers) aligns with the strengths of AI-era Make. The RBV argument reinforces this: these applications are candidates for competitive differentiation that standardized enterprise software packages cannot provide.

\paragraph{Category 3: Regulated standard applications} (medium complexity, low specificity, high compliance). Examples include HR systems in regulated industries, financial reporting tools, and healthcare record systems. AI-induced MoB shift: \emph{low}. High compliance exposure creates governance costs (conformity assessment, audit documentation, and liability attribution) that structurally favor certified vendors with proven compliance track records, whether they supply software in the cloud or as on-premise deployments. The cost reduction from AI is insufficient to offset the compliance risk premium of in-house development.

\paragraph{Category 4: Mission-critical systems of record} (high complexity, medium-high specificity, high compliance). Examples include ERP core modules, core banking systems, and supply chain platforms. AI-induced MoB shift: \emph{very low}. The combination of high complexity, deep integration requirements (cf.\ the multi-system integration limitation identified in Section~2.2), and stringent compliance obligations preserves the case for established vendors. Make via AI tools may be viable for peripheral modules or extensions, but not for the core system in the near term.

Table~\ref{tab:typology} summarizes the four categories and their AI-induced MoB sensitivity.

\begin{table}[ht]
\centering
\caption{Application typology by MoB-AI sensitivity.}
\label{tab:typology}
\small
%\begin{tabular}{p{3.2cm}cccp{2.8cm}}
\begin{tabular}{p{3.2cm}ccc>{\raggedright\arraybackslash}p{2.8cm}}
\hline
\textbf{Category} & \textbf{Complexity} & \textbf{Specificity} & \textbf{Compliance} & \textbf{AI-Induced MoB Shift} \\
\hline
Commodity utilities & Low & Low & Low & High $\rightarrow$ Make \\
Differentiating custom & Low-Med & High & Low-Med & High $\rightarrow$ Make \\
Regulated standard & Medium & Low & High & Low $\rightarrow$ Buy \\
Mission-critical SoR & High & Med-High & High & Very low $\rightarrow$ Buy \\
\hline
\end{tabular}
\end{table}

\subsection{Revised Decision Logic}

Building on the typology and the factor analysis, we propose a two-stage decision model for the AI era:

\paragraph{Stage 1: Application classification.} The firm classifies each application (or application component) into one of the four typology categories, based on its complexity, domain specificity, and compliance exposure. This classification determines the baseline MoB recommendation and the relative weighting of the seven decision factors.

\paragraph{Stage 2: Factor weighting and capability assessment.} The firm applies the AI-adjusted factor assessments from Section~3, modulated by two firm-specific variables: (a)~the firm's current \emph{organizational AI capability readiness}, the threshold variable identified in Section~3.7, and (b)~the firm's specific regulatory context. Below a minimum level of AI development capability, Make via AI tools is not viable regardless of how favorable the application-level factors may be.

The resulting decision heuristics per category are:

\begin{itemize}
    \item \textbf{Commodity utilities}: Default to Make. Evaluate Buy only where ecosystem integrations provide strong network value or where the firm's AI capability is below the viability threshold.
    \item \textbf{Differentiating custom applications}: Evaluate Make first. Revert to Buy only where time pressure is acute, capability gaps are prohibitive, or the application requires integration depth beyond current AI capabilities.
    \item \textbf{Regulated standard applications}: Default to Buy from certified vendors. Revisit periodically as AI compliance tooling matures and as the regulatory framework for AI-generated software clarifies.
    \item \textbf{Mission-critical systems of record}: Retain Buy as the primary option. Consider Make selectively for peripheral modules, extensions, or integration layers where the core system's integrity is not at risk.
\end{itemize}

A critical implication of this framework is that \emph{organizational AI capability readiness} functions as a gateway condition. Firms below the capability threshold face a preliminary strategic decision: invest in building AI development capability (a prerequisite for Make in any category) or accept continued dependence on external procurement. This capability investment decision is itself a strategic choice with long-term competitive implications, as the analysis in Section~3.7 suggests that AI orchestration capability may satisfy the RBV's criteria for a sustainable competitive resource.

\section{Discussion}

\subsection{Implications for IS Strategy}

The framework developed in this paper suggests three strategic implications for IS management. First, firms should conduct a \emph{portfolio-level reassessment} of their application landscape against the typology proposed in Section~4.2, systematically identifying candidates for in-house Make. This reassessment is most likely to yield actionable results in the commodity utility and differentiating custom application categories, where the economics of AI-assisted development are most compelling.

Second, the analysis points to an \emph{unbundling dynamic}: AI enables selective replacement of individual modules or application components rather than wholesale replacement of entire vendor platforms. This incremental, lower-risk strategy contrasts with the dramatic SaaSocalypse narrative and is more consistent with the nuanced, application-dependent pattern of MoB shifts identified in Section~3. Firms may, for example, retain a core enterprise application platform for its compliance infrastructure and ecosystem integrations while building custom extensions or peripheral applications in-house. From an IT strategy perspective, the traditional ``buy'' decision has always entailed a secondary choice of sourcing strategy \citep{LightEtAl2001, Bradley2009}. A \emph{best-of-breed} approach optimizes fit to business requirements by sourcing from specialized vendors, but incurs significant integration costs from a heterogeneous vendor landscape. A \emph{best-of-suite} approach trades configurability for integration simplicity: fit to specific requirements is typically lower, but components interoperate out of the box, reducing integration overhead. AI-augmented make may narrow this long-standing tension. Where selective, module-level custom development becomes economically viable, firms can combine vendor-provided foundations, e.g., data models, core process templates, and compliance infrastructure from one or more platform vendors, with custom-developed modules and components that deliver high differentiation. This hybrid architecture collapses the best-of-breed versus best-of-suite binary into a more granular, capability-driven decision: which components are sufficiently standard that a vendor provides them at lower cost and risk, and which are sufficiently differentiating that AI-assisted in-house development is justified.

Third, \emph{capability investment} emerges as a prerequisite for realizing the potential of Make in the AI era. The organizational AI capability readiness threshold identified in Section~3.7 implies that firms wishing to exercise the Make option must invest deliberately in AI orchestration skills, governance structures, and quality assurance practices. This investment is itself a strategic decision with long-term implications, as the RBV analysis suggests that AI development capability may become a source of sustained competitive advantage.

\subsection{Practical Challenges}

Several practical challenges constrain the adoption of Make in the AI era. The \emph{regulatory landscape} imposes obligations that differ depending on the nature of the application being built: firms must ensure that custom-developed software complies with applicable laws and sectoral regulations, and bear full legal responsibility for doing so, as analyzed in detail in Section~3.6.

\emph{Data privacy and sovereignty} present a related constraint. The AI development pipeline itself becomes a potential data exposure vector when proprietary code, business logic, or customer data are transmitted to third-party model providers. Firms in jurisdictions with strict data sovereignty requirements may face constraints on which AI infrastructure they can use, limiting the practical scope of Make in the AI era.

\emph{Software quality and security} risks are qualitatively different for AI-generated code. Liability allocation for defects in AI-generated artifacts remains legally unsettled, and security audit practices for AI-assisted development are nascent. The ``last mile'' quality risks identified in Section~3.6 require dedicated governance processes that most organizations have not yet established.

\emph{Technical debt} may accumulate differently in AI-generated codebases. Without disciplined governance, AI-generated code may exhibit structural entropy---inconsistent patterns, redundant logic, and opaque design decisions---that increases long-run maintenance costs. The OPEX implications may be less favorable than initial projections suggest if technical debt is not actively managed.

Finally, \emph{organizational change} presents a binding constraint. Reskilling existing IT staff, overcoming resistance to AI-augmented workflows, and establishing governance maturity for AI-generated artifacts require sustained organizational effort that extends well beyond the technical implementation.

\subsection{Competitive Dynamics: The Co-Evolution of Make and Buy}

A key limitation of static MoB analysis is that it treats the ``make'' and ``buy'' options as independent. In practice, AI transforms both sides simultaneously, creating co-evolutionary dynamics that will shape the enterprise software landscape over time. In the short term, the most visible effect is the expansion of the Make frontier into application categories that were previously uneconomical to build in-house. This expansion is concentrated in commodity utilities and differentiating custom applications, where the analysis in Section~3 identifies the strongest net shifts toward ``make.'' In the medium term, enterprise application vendors will respond strategically, embedding AI more deeply, expanding platform ecosystems, and potentially restructuring pricing models to defend their installed bases. The competitive dynamic is not a one-time disruption but an ongoing co-evolution in which AI capabilities on both sides advance in parallel.

The likely outcome is a new market segmentation: Enterprise application vendors retain their strongest position in regulated, complex, and compliance-intensive categories (Categories 3 and 4 of the typology), while Make captures an increasing share of specific, low-compliance applications (Categories 1 and 2). This segmentation is not static; as AI capabilities mature and compliance tooling develops, the boundaries between categories will shift, warranting periodic reassessment of the application portfolio. This projection represents a hypothesis for empirical investigation rather than a definitive prediction; the actual trajectory will depend on factors---including vendor strategic responses, regulatory developments, and the pace of AI capability advancement---that are inherently difficult to forecast.

A related consideration concerns the trajectory of Make's hybrid governance form over time. As AI development tools become ubiquitous and their governance implications are absorbed into standard organizational practices, the hybrid characteristics of AI-era Make (variable-cost AI production inputs, information asymmetry with AI providers, and non-deterministic production processes) may gradually be normalized into a redefined baseline for in-house development. This mirrors earlier technological transitions: cloud computing and open-source tooling also reshaped what ``Make'' entails without permanently requiring separate analytical categories. The structural properties identified in Section~4.1 suggest, however, that this normalization will extend over a substantial period. During this transition, the governance, capability, and strategic implications of AI-era Make warrant the distinct treatment provided here. The framework is thus best understood as capturing a strategically consequential transition whose implications for IS strategy are significant regardless of whether the hybrid character of Make eventually becomes the unremarkable default.

\subsection{Limitations}

This study has several limitations that should inform interpretation and future research. One key limitation is that the presented framework is conceptual. While grounded in established IS theories and informed by current AI capabilities, it has not been empirically validated across industries, firm sizes, or regulatory contexts. The factor assessments and typology categories represent theoretically derived propositions that require empirical testing. Given that AI-augmented development at enterprise scale is still in its early stages of adoption, it would be premature to subject the framework to rigorous empirical validation at this point: the organizational practices, governance structures, and cost data needed for meaningful testing are not yet widely available. Even a cross-sectional study would likely capture only initial adoption patterns before firms have had sufficient time to internalize the new capabilities and adapt their sourcing strategies accordingly. A longitudinal research design therefore seems necessary to trace how make-or-buy decisions evolve as AI capabilities mature and organizational experience accumulates.

Another limiting aspect is that AI capabilities are advancing rapidly. The specific capability boundaries, cost structures, and quality characteristics described in Section~2.2 reflect the state of the art as of early 2026. As agentic systems improve, the thresholds and factor magnitudes identified in Section~3 will shift, potentially altering the typology boundaries and decision heuristics. Furthermore, the framework does not yet model the competitive dynamics between AI-assisted in-house development and AI-accelerated enterprise software vendors over time. While Section~5.3 discusses these co-evolutionary dynamics qualitatively, a formal model of the strategic interaction would further strengthen the framework's predictive power. Longitudinal studies tracking actual insourcing decisions and their outcomes are needed to validate and refine the proposed decision logic. In addition, the factor assessments and typology classifications reflect analytical judgment informed by current evidence rather than empirically measured magnitudes. Different weighting of the same factors could yield different recommendations, particularly for applications that fall near category boundaries.

Finally, the framework treats ``Buy'' as a single option and does not fully distinguish the procurement sub-options it encompasses. SaaS subscriptions, private cloud deployments, and traditional on-premise installations from established enterprise application vendors are distinct delivery models with different cost structures, lock-in profiles, and compliance implications. It also does not specifically address open-source software as a distinct procurement alternative, which may offer a different trade-off profile. Similarly, the regulatory analysis centers on the European Union's legal framework; firms operating under different regulatory regimes may face different compliance calculi.

\section{Conclusion}

This paper has examined how advances in generative AI, particularly agentic coding systems, alter the Make-or-Buy decision for enterprise applications. The central finding is that AI-assisted software development does not invalidate canonical MoB frameworks grounded in TCE and RBV; rather, it recalibrates their key variables, most significantly by lowering the cost floor for ``make'' and by compressing development timelines. The rational strategic response is not the wholesale abandonment of packaged enterprise applications that the SaaSocalypse narrative implies, but a selective, portfolio-level reassessment guided by application type, compliance context, and organizational capability.

The paper's primary theoretical contribution is the analysis of how AI transforms the governance structure of the Make option itself, shifting it from Williamson's pure hierarchy to a hybrid form that combines code ownership and internal control with dependence on external AI infrastructure. This transformation is characterized by qualitatively different capability requirements, cost dynamics, and governance structures relative to pre-AI in-house development. The proposed application typology provides a structured basis for determining where this transformed Make option is most compelling: commodity utilities and differentiating custom applications emerge as the categories with the highest AI-induced MoB shift, while regulated standard applications and mission-critical systems of record remain predominantly in the ``buy'' domain.

Three managerial implications follow. First, firms should re-evaluate their application portfolio using the proposed typology to identify candidates for in-house Make. Second, investment in AI development capability (e.g., prompt engineering, agent orchestration, and AI governance) should be treated as a strategic asset, not merely as a cost-reduction measure, since this capability may itself become a source of sustained competitive advantage. Third, firms should not underestimate the operational complexity, compliance exposure, and technical debt risks associated with AI-generated solutions; the 60--80\% operational cost share of the software lifecycle does not disappear simply because the development share shrinks.

Future research should pursue several directions: (1)~empirical validation of the framework through case studies and surveys of AI-driven insourcing decisions; (2)~updated TCO models that incorporate AI infrastructure and governance costs; (3)~longitudinal analyses of organizational AI capability development; and (4)~formal modeling of the competitive dynamics between AI-enabled in-house development and AI-accelerated SaaS vendors.

\bibliographystyle{unsrtnat}
\bibliography{references}  %%% Uncomment this line and comment out the ``thebibliography'' section below to use the external .bib file (using bibtex) .

%%% Uncomment this section and comment out the \bibliography{references} line above to use inline references.
% \begin{thebibliography}{1}

% 	\bibitem{kour2014real}
% 	George Kour and Raid Saabne.
% 	\newblock Real-time segmentation of on-line handwritten arabic script.
% 	\newblock In {\em Frontiers in Handwriting Recognition (ICFHR), 2014 14th
% 			International Conference on}, pages 417--422. IEEE, 2014.

% 	\bibitem{kour2014fast}
% 	George Kour and Raid Saabne.
% 	\newblock Fast classification of handwritten on-line arabic characters.
% 	\newblock In {\em Soft Computing and Pattern Recognition (SoCPaR), 2014 6th
% 			International Conference of}, pages 312--318. IEEE, 2014.

% 	\bibitem{keshet2016prediction}
% 	Keshet, Renato, Alina Maor, and George Kour.
% 	\newblock Prediction-Based, Prioritized Market-Share Insight Extraction.
% 	\newblock In {\em Advanced Data Mining and Applications (ADMA), 2016 12th International 
%                       Conference of}, pages 81--94,2016.

% \end{thebibliography}

\end{document}